\documentclass[dvips]{arxstspdf}
\usepackage{flushend}
\usepackage{stfloats}
\usepackage{graphicx}

\volume{26}
\issue{1}
\pubyear{2011}
\firstpage{1}
\lastpage{9}
\doi{10.1214/10-STS337}

\makeatletter
\newproclaim{Frequentist assumptions}{Frequentist assumptions}
\newproclaim{Frequentist interpretation of confidence interval}{Frequentist interpretation of confidence interval}
\newproclaim{Pragmatic interpretation of confidence interval}{Pragmatic interpretation of confidence\break interval}
\newproclaim{Bayesian assumptions}{Bayesian assumptions}
\newproclaim{Bayesian interpretation of posterior interval}{Bayesian interpretation of posterior interval}
\newproclaim{Pragmatic interpretation of posterior interval}{Pragmatic interpretation of posterior interval}
\makeatother

\begin{document}
\begin{frontmatter}

\title{Statistical Inference: The Big Picture{\thanksref{T1}}}
\relateddoi{T1}{Discussed in \doi{10.1214/11-STS337C},
\doi{10.1214/11-STS337A}, \doi
{10.1214/11-STS337D} and \doi{10.1214/11-STS337B}; rejoinder at \doi{10.1214/11-STS337REJ}.}
\runtitle{Statistical Inference}

\begin{aug}
\author[a]{\fnms{Robert E.} \snm{Kass}\corref{}\ead[label=e1]{kass@stat.cmu.edu}}
\runauthor{R. E. Kass}

\affiliation{Carnegie Mellon University}

\address[a]{Robert E. Kass is Professor,
Department of Statistics, Center
  for the Neural Basis of
  Cognition and Machine Learning Department, Carnegie Mellon
  University, Pittsburgh, Pennsylvania 15213, USA \printead{e1}.}

\end{aug}

\begin{abstract}
Statistics
has moved beyond the frequentist-Bayesian controversies
of the past. Where does this leave our ability to
interpret results? I suggest that a
philosophy compatible with  statistical practice,
labeled here \textit{statistical pragmatism},
serves as a foundation for inference. Statistical
pragmatism is inclusive and emphasizes the assumptions
that connect statistical models with observed data.
I argue that introductory courses often
mischaracterize the process of statistical inference
and I propose an alternative ``big picture'' depiction.
\end{abstract}


\end{frontmatter}

\section{Introduction}

The protracted battle for the foundations of statistics, joined
vociferously by Fisher, Jeffreys, Neyman, Savage and many disciples,
has been deeply illuminating, but it has left statistics without a
philosophy that matches contemporary attitudes. Because each camp took
as its goal exclusive ownership of inference, each was doomed to
failure.  We have all, or nearly all, moved past these old debates,
yet our textbook explanations have not caught up with the eclecticism
of statistical practice.

The difficulties go both ways.  Bayesians have denied the utility of
confidence and statistical significance, attempting to sweep aside the
obvious success of these concepts in applied work. Meanwhile, for
their part, frequentists have ignored the possibility of inference
about unique events despite their ubiquitous occurrence throughout
science.  Furthermore, interpretations of posterior probability in
terms of subjective belief, or confidence in terms of long-run
frequency, give students a limited and sometimes confusing view of the
nature of statistical inference. When used to introduce
the expression of uncertainty based on a random sample,
these caricatures forfeit an opportunity to articulate a fundamental
attitude of statistical practice.

Most modern practitioners have, I think, an open-minded view about
alternative modes of inference, but are acutely aware of theoretical
assumptions and the many ways they may be mistaken. I would suggest
that it makes more sense to place in the center of our logical
framework the match or mismatch of theoretical assumptions with the
real world of data. This, it seems to me, is the common ground that
Bayesian and frequentist statistics share; it is more fundamental than
either paradigm taken separately; and as we strive to foster
widespread understanding of statistical reasoning, it is more
important for beginning students to appreciate the role of theoretical
assumptions than for them to recite correctly
the
long-run interpretation of confidence intervals. With the hope
of prodding our discipline to right a~lingering imbalance,
I attempt here to describe the dominant contemporary philosophy of
statistics.

\section{Statistical Pragmatism}\label{sec2}

\begin{figure}

\includegraphics{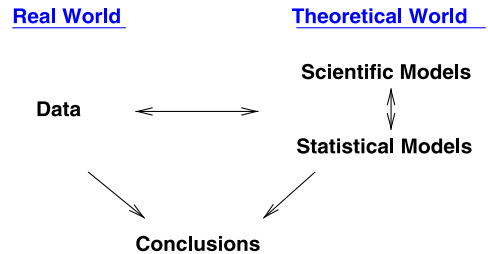}

\caption{The big picture of
statistical inference. Statistical
procedures are abstractly defined in terms of  mathematics but are
used, in conjunction with scientific models and methods,
to explain\break observable phenomena. This picture emphasizes
the hypothetical link between variation in data and its description
using statistical models.}\label{fig1}
\end{figure}

I propose to call this modern philosophy
\textit{statistical pragmatism}. I think it is based on the
following attitudes:
\begin{enumerate}
\item Confidence, statistical significance, and posterior probability
are all valuable inferential tools.
\item Simple chance
situations, where counting arguments may be based on symmetries
that generate equally likely outcomes
(six faces on a fair die; 52 cards in a shuffled deck), supply basic
intuitions about probability. Probability may be built up
to important but less immediately intuitive
situations using abstract mathematics, much
the way real numbers are defined abstractly based on intuitions coming
from fractions. Probability is usefully calibrated in terms of fair bets:
another way
to say the probability of rolling a 3 with a fair die is $1/6$ is that
5 to 1 odds against rolling a 3 would be a fair bet.

\item Long-run frequencies are important mathematically, interpretively,
and pedagogically. However, it
is possible to assign probabilities to unique events,
including rolling a 3 with a fair die or having
a confidence interval cover the true mean, without considering
long-run frequency. Long-run frequencies may be regarded
as consequences of the law of large numbers rather than as part
of the definition of probability or confidence.
\item Similarly, the subjective interpretation of posterior
probability is important as a way of understanding
Bayesian inference, but it is not fundamental to its use:
in reporting a 95\% posterior interval one
need not make a statement such as, ``My
personal probability of this interval covering the mean is
0.95.''

\item Statistical inferences of all kinds use statistical
models, which embody theoretical assumptions. As illustrated
in Figure \ref{fig1}, like scientific models,
statistical models exist in an abstract framework; to distinguish
this framework from the real world inhabited by data we may
call it a ``theoretical world.''
Random variables, confidence intervals, and posterior probabilities
all live in this theoretical world. When
we use a statistical model to make a statistical inference
we implicitly assert that the variation exhibited by data
is captured reasonably well by the statistical model, so that
the theoretical world corresponds reasonably well to the real
world. Conclusions are drawn by applying a statistical
inference technique, which is a theoretical construct, to some
real data.
Figure \ref{fig1} depicts the conclusions
as straddling the theoretical and real worlds.
Statistical inferences may have implications for the real
world of new observable phenomena, but in scientific
contexts, conclusions most often concern scientific models (or theories),
so that their ``real world'' implications (involving new data)
are somewhat indirect (the new data will involve new and different
experiments).
\end{enumerate}

The statistical models in Figure \ref{fig1} could involve large
function spaces or other relatively weak probabilistic assumptions.
Careful consideration of the connection between models and data is a
core component of both the art of statistical practice and the science
of statistical methodology.  The purpose of Figure \ref{fig1} is to
shift the grounds for discussion.

Note, in particular, that data should not be confused with
random variables.
Random variables live in the theoretical world.
When we say things like, ``Let us assume the data are normally distributed''
and we proceed to make a statistical inference, we do not need
to take these words literally as asserting that
the data form a random sample. Instead, this kind of language is a
convenient and familiar shorthand for the much weaker
assertion that, for our specified purposes,
the variability of the data is adequately consistent
with variability that would occur in a random sample.
This linguistic amenity is used routinely in
both frequentist and Bayesian frameworks.
Historically, the distinction
between data and random variables, the match of the model to the data,
was set aside, to be treated as a separate topic apart from the
foundations of inference.  But once the data themselves were
considered random variables, the frequentist-Bayesian debate moved
into the theoretical world: it became a debate about the best way to
reason from random variables to inferences about parameters.  This was
consistent with developments elsewhere. In other parts of science, the
distinction between quantities to be measured\vadjust{\eject} and their theoretical
counterparts within a mathematical theory can be relegated to a
different sub\-ject---to a theory of errors. In statistics, we do not have
that luxury, and it seems to me important, from a pragmatic viewpoint,
to bring to center stage the identification of models with data.
The purpose of doing so is that it provides different interpretations
of both frequentist and Bayesian inference, interpretations
which, I believe, are closer to the attitude of
modern statistical practitioners.
\begin{figure*}
\begin{tabular}{@{}c@{}}

\includegraphics{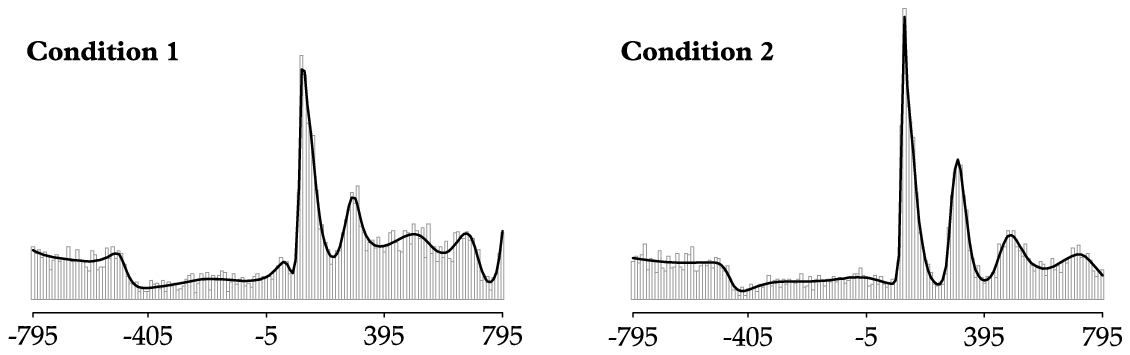}
\\
\scriptsize{(A)}\\

\includegraphics{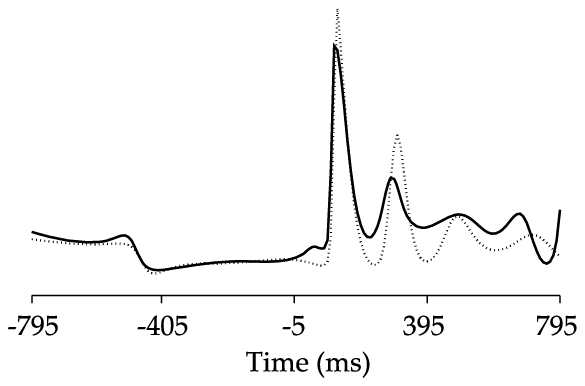}
\\
\scriptsize{(B)}
\end{tabular}
\caption{\textup{(A)} BARS fits to a pair of peri-stimulus time histograms
displaying neural firing rate of a particular neuron
under two alternative experimental conditions.
\textup{(B)} The two BARS fits are overlaid for ease of comparison.}\label{fig:rollenhagen}
\end{figure*}

A familiar practical situation where these issues arise is binary
regression. A classic example comes from a psychophysical experiment
conducted by\break Hecht, Schlaer and Pirenne (\citeyear{HSP1942}), who
investigated the
sensitivity of the human visual system by
constructing an apparatus that would emit flashes of
light at very low intensity in a darkened room. Those authors presented light
of varying intensities repeatedly to several subjects and determined,
for each intensity, the proportion of times each subject would respond
that he or she had seen a flash of light. For each subject the
resulting data are repeated binary observations (``yes'' perceived versus
``no'' did not perceive) at each of many intensities and, these days,
the standard statistical tool to analyze such data is logistic
regression.  We might, for instance, use maximum likelihood to find a
95\% confidence interval for the intensity of light at which the
subject would report perception with probability $p=0.5$.  Because the
data reported by Hecht et al. involved fairly large samples, we
would obtain essentially the same answer if instead we applied
Bayesian methods to get an interval having 95\% posterior probability.
But how should such an interval be interpreted?

A more recent example comes from
DiMatteo, Genovese and Kass (\citeyear{DGK2001}), who illustrated
a new nonparametric regression method called
Bayesian adaptive regression splines (BARS) by
analyzing neural firing rate data from
inferotemporal cortex of a macaque monkey.
The data came from
a study ultimately reported by Rollenhagen and Olson (\citeyear{RO2005}),
which investigated the differential response of individual
neurons under two experimental
conditions. Figure \ref{fig:rollenhagen} displays BARS fits
under the two conditions.
One way to quantify the discrepancy between the fits
is to estimate the drop in firing
  rate from peak (the maximal firing rate) to the trough
  immediately following the peak in each condition.
Let us call these peak minus trough
  differences, under the two conditions, $\phi^1$ and $\phi^2$.
Using BARS, DiMatteo, Genovese and
  Kass reported a posterior mean
of $\hat \phi^1 - \hat \phi^2 = 50.0$ with
posterior standard deviation $(\pm 20.8)$.
In follow-up work,
Wallstrom, Liebner and Kass (\citeyear{WLK2008}) reported very good frequentist
coverage probability of  95\% posterior probability intervals
based on BARS
for similar quantities under simulation conditions chosen to mimic
such experimental data. Thus, a BARS-based posterior interval
could be considered
from either a Bayesian or frequentist point of view. Again we
may ask how such an inferential interval should be interpreted.

\section{Interpretations}

Statistical pragmatism involves mildly altered interpretations
of frequentist and Bayesian inference. For definiteness I will discuss
the paradigm case of confidence and posterior intervals for a normal mean
based on a sample of size $n$, with the standard deviation being known.
Suppose that we have $n=49$ observations that have
a sample mean equal to $10.2$.
\begin{Frequentist assumptions*}
Suppose $X_1,X_2,\break\ldots,X_n$ are i.i.d. random variables
from a normal distribution with mean $\mu$ and standard deviation
$\sigma=1$. In other words, suppose
$X_1,X_2,\ldots,X_n$ form a random
sample from a $N(\mu,1)$ distribution.
\end{Frequentist assumptions*}

Noting that $\bar{x}=10.2$ and $\sqrt{49}=7$
we define the inferential interval
\[
I = \bigl(10.2-\tfrac{2}{7},
10.2+\tfrac{2}{7}\bigr).
\]
The interval $I$ may be regarded as a 95\% confidence interval.
I now contrast the standard frequentist interpretation with
the pragmatic interepretation.
\begin{Frequentist interpretation of confidence interval*}
Under the assumptions above, if
we we\-re to draw infinitely many random samples from a
$N(\mu,1)$ distribution, 95\% of
the corresponding confidence intervals
$(\bar{X}-\frac{2}{7},\bar{X}+\frac{2}{7})$
would cover $\mu$.
\end{Frequentist interpretation of confidence interval*}
\begin{Pragmatic interpretation of confidence interval*}
If we were to draw a random sample according to the
assumptions\vspace*{1pt}
above, the resulting
confidence interval $(\bar{X}-\frac{2}{7},\bar{X}+\frac{2}{7})$
would have probability 0.95 of covering $\mu$.
Because the random sample
lives in the theoretical world, this is a theoretical statement.
Nonetheless, substituting
%
\begin{equation}\label{bar1}
\bar{X} =\bar{x}
\end{equation}
together with
%
\begin{equation}\label{bar2}
\bar{x} =10.2
\end{equation}
we obtain the interval $I$, and
are able to draw useful conclusions
as long as our theoretical world is aligned well with the
real world that produced the data.
\end{Pragmatic interpretation of confidence interval*}

The main point here is that we do not need a long-run interpretation
of probability, but we do have to be reminded that the unique-event
probability of 0.95 remains a theoretical statement because it applies
to random variables rather than data. Let us turn
to the Bayesian case.
\begin{Bayesian assumptions*}
Suppose $X_1,X_2,\break\ldots,X_n$ form a random
sample from a $N(\mu,1)$ distribution and the prior distribution
of $\mu$ is $N(\mu_0,\tau^2)$, with $\tau^2 \gg \frac{1}{49}$ and
$49\tau^2 \gg |\mu_0|$.
\end{Bayesian assumptions*}

The posterior distribution of $\mu$ is normal, the
posterior mean becomes
\[
\bar{\mu} = \frac{\tau^2}{1/49+\tau^2}10.2 +
\frac{1/49}{1/49+\tau^2}\mu_0
\]
and the posterior variance is
\[
v = \biggl(49 +\frac{1}{\tau^2}\biggr)^{-1}
\]
but because $\tau^2 \gg \frac{1}{49}$
and $49\tau^2 \gg |\mu_0|$ we have
\[
\bar{\mu} \approx 10.2
\]
and
\[
v \approx \frac{1}{49}.
\]
Therefore, the inferential interval $I$ defined above
has posterior probability 0.95.

\begin{Bayesian interpretation of posterior interval*}
Under the assumptions above,
the probability that $\mu$ is in the interval $I$ is 0.95.
\end{Bayesian interpretation of posterior interval*}
\begin{Pragmatic interpretation of posterior interval*}
If the data were
a random sample for\break which~(\ref{bar2}) holds, that is,
$\bar{x}=10.2$, and if the as\-sumptions above were to hold,
then the probability that $\mu$ is
in the interval $I$ would be 0.95.  This refers to a hypothetical value
$\bar{x}$ of the random variable $\bar{X}$, and because $\bar{X}$
lives in the theoretical world the statement remains theoretical.
Nonetheless, we
are able to draw useful conclusions from the data
as long as
our theoretical world is aligned well with the real world that
produced the data.
\end{Pragmatic interpretation of posterior interval*}

Here, although the Bayesian approach escapes
the indirectness of confidence within the theoretical\break world,
it cannot escape it in the world of data analysis because there
remains the additional layer of identifying data with random variables.
According to the pragmatic interpretation, the posterior is not,
literally, a
statement about the way the observed data relate to the unknown
parameter $\mu$ because those objects live in different
worlds. The language of Bayesian inference, like the language of
frequentist inference, takes a convenient shortcut by
blurring the
distinction between data and random variables.

The commonality between frequentist and Baye\-sian
inferences is the use of theoretical assumptions, together
with a \textit{subjunctive} statement. In both approaches
a statistical model is introduced---in the Bayesian case
the prior distributions become part of what I am here calling
the model---and we may say that the
inference is based on what \textit{would} happen
if the data  \textit{were} to be random variables
distributed according to the statistical model.
This modeling assumption would
be reasonable if the model
\textit{were} to describe accurately
the variation in the data.

\section{Implications for Teaching}\label{sec4}

\begin{figure}[b]

\includegraphics{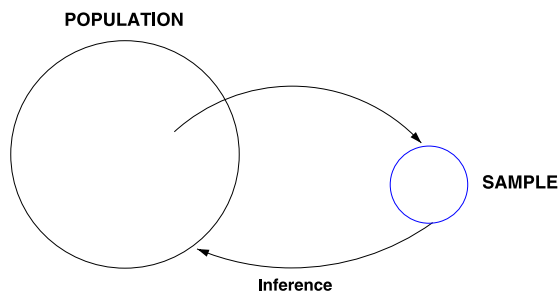}

\caption{The big picture of statistical inference
according to the standard conception.
Here, a random sample is pictured as a
sample from a finite population.}\label{fig2}
\end{figure}

It is important for students in introductory statistics courses
to see the subject as a coherent, principled whole.
Instructors, and textbook authors, may try to help
by providing some notion of a ``big picture.'' Often this
is done literally, with an illustration such as
Figure~\ref{fig2} (e.g., Lovett, Meyer and Thille, \citeyear{LMT2008}).
This kind of illustration can be extremely useful if referenced
repeatedly throughout a course.

Figure \ref{fig2} represents
a standard story about statistical inference.
Fisher introduced the idea of a~random sample drawn from a~%
hypothetical infinite population, and Neyman and Pearson's work
encouraged subsequent mathematical statisticians to
drop the word ``hypothetical'' and instead
describe statistical inference as analogous to simple random sampling from
a finite population. This is the concept that
Figure \ref{fig2} tries to get across.
My complaint is that it is not a good general description
of statistical inference, and my claim is that Figure~\ref{fig1} is
more accurate.
For instance, in the psychophysical example of Hecht, Schlaer and Pirenne discussed
in Section \ref{sec2}, there is no population of ``yes'' or ``no''
replies from which a random sample is drawn. We do not need to
struggle to make an analogy with a simple random sample. Furthermore, any
thoughts along these lines may draw attention away from
the most important theoretical assumptions, such as independence among
the responses. Figure \ref{fig1} is supposed to remind students to
look for the important assumptions, and ask whether they describe
the variation in the data reasonably accurately.

One of the reasons the population and sample picture in Figure \ref{fig2}
is so attractive pedagogically is that it reinforces the fundamental
distinction between parameters and statistics through the terms
\textit{population mean} and \textit{sample mean}. To my way of thinking,
this terminology, inherited from Fisher, is unfortunate.
Instead of ``population mean'' I would much prefer
\textit{theoretical mean}, because it captures better the notion
that a theoretical distribution is being introduced, a notion that is
reinforced by Figure \ref{fig1}.

I have found Figure \ref{fig1} helpful in teaching
basic statistics. For instance, when talking about
random variables I like to begin with
a set of data, where variation is displayed in a histogram,
and then say that probability may be used to describe
such variation. I then tell the students we must introduce mathematical
objects called random variables, and in defining them and applying
the concept to the data at hand, I immediately acknowledge
that this is an abstraction, while also stating that---as the
students will see repeatedly in many examples---it can
be an extraordinarily useful abstraction
whenever the theoretical
world of random variables is aligned  well with
the real world of the data.

I have also used Figure \ref{fig1} in my classes
when describing attitudes toward data analysis
that statistical training aims to instill.
Specifically, I define statistical thinking, as
in the article by Brown and Kass (\citeyear{BK2009}),  to involve two principles:
\begin{enumerate}
\item
Statistical models  of regularity and variability
in data may be used to express knowledge and uncertainty
about a signal in the presence of noise,
via inductive reasoning.
\item Statistical methods may be analyzed to determine how
well they are likely to perform.
\end{enumerate}
Principle 1 identifies the source of statistical inference to be
the hypothesized link between data and statistical models.
In explaining, I
explicitly distinguish
the use of probability to describe variation and to express
knowledge. A probabilistic description of variation would
be
``The probability of rolling a 3 with a fair die is $1/6$''
while an expression of knowledge would be
``I'm 90\% sure
the capital of Wyoming is Cheyenne.''
These two sorts of statements, which use probability in different
ways, are sometimes considered to involve two
different kinds of probability,
which have been called
``aleatory probability'' and ``epistemic probability.''
Bayesians mer\-ge these,
applying the laws of probability to
go from quantitative
description to quantified belief, but in every
form of statistical inference aleatory probability
is used, somehow, to make epistemic statements.
This is Principle 1. Principle 2 is that the same sorts
of statistical models may be used to evaluate statistical
procedures---though in the classroom I also explain that performance
of procedures is usually investigated under varying circumstances.
\begin{figure*}

\includegraphics{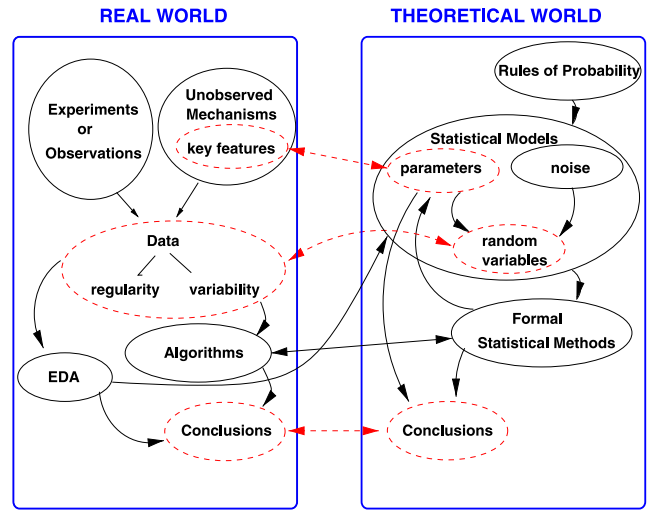}

\caption{A more elaborate big picture, reflecting in
greater detail the process of statistical inference. As in
Figure \protect\ref{fig1}, there is a hypothetical link between data
and statistical models but here the data are connected more specifically
to their representation as random variables.}\label{bigpic5}
\end{figure*}

For somewhat more advanced audiences it is
possible to elaborate, describing in more detail
the process trained statisticians follow when reasoning
from data. A big picture of the overall process is given
in Figure~\ref{bigpic5}. That figure indicates the hypothetical connection
between data and random variables, between key features of unobserved
mechanisms and parameters, and between real-world and theoretical
conclusions. It further indicates that data display both regularity
(which is often described in theoretical terms as  a ``signal,''
sometimes conforming to simple mathematical descriptions or ``laws'')
and unexplained variability, which is usually taken to be ``noise.''
The figure also includes
the components exploratory data analysis---EDA---and algorithms,
but the main message of
Figure~\ref{bigpic5}, given by the labels of the two big boxes,
is the same as that in
Figure \ref{fig1}.

\section{Discussion}

According to my understanding, laid out above, statistical pragmatism
has two main features: it is eclectic and it emphasizes the
assumptions that connect statistical models with observed data.  The
pragmatic view acknowledges that both sides of the
fre\-quentist-Bayesian debate made important points.\break  Bayesians scoffed
at the artificiality in using sampling from a finite population to
motivate all of inference, and in using long-run behavior to define
characteristics of procedures. Within the theoretical world, posterior
probabilities are more direct, and therefore seemed to offer much
stronger inferences. Frequentists bristled, pointing to the
subjectivity of prior distributions. Bayesians responded by
treating subjectivity as a virtue on the grounds that all inferences
are subjective yet, while there is a kernel of truth in this
observation---we are all human beings, making our own
judgments---subjectivism was never satisfying as a logical framework:
an important purpose of the scientific enterprise is to go beyond
personal decision-making.  Nonetheless, from a~%
pragmatic perspective, while the selection
of prior probabilities is important, their use is not
so problematic as to disqualify Bayesian methods, and in looking back
on history the
introduction of prior distributions may not have been the central
bothersome issue it was made out to be. Instead, it seems to me, the
really troubling point for frequentists has been the Bayesian claim to
a philosophical high ground, where compelling inferences could be
delivered at negligible logical cost. Frequentists have always felt
that no such thing should be possible. The difficulty begins
not with the introduction of prior distributions but with the gap
between models and data, which is neither frequentist nor Bayesian.
Statistical pragmatism avoids this irritation by acknowledging
explicitly the tenuous connection between the real and theoretical
worlds. As a result, its inferences are necessarily subjunctive. We
speak of what \textit{would} be inferred if our assumptions \textit{were} to
hold.  The inferential bridge is traversed, by both frequentist and
Bayesian methods, when we act \textit{as if} the data \textit{were}
generated by random variables. In the normal mean example discussed in
Section \ref{sec4}, the key step involves the conjunction of the two equations
(\ref{bar1}) and (\ref{bar2}).  Strictly speaking, according to
statistical pragmatism, equation (\ref{bar1}) lives in the theoretical
world while equation (\ref{bar2}) lives in the real world; the bridge
is built by allowing $\bar{x}$ to refer to \textit{both} the theoretical
value of the random variable and the observed data value.

In pondering the nature of statistical
inference I am, like others, guided
partly by past and present sages (for an overview
see Barnett, \citeyear{Barnett1999}), but also by my own experience
and by watching many colleagues in action.
Many of the  sharpest and most vicious Bayes-frequentist debates
took place during the dominance of pure theory
in academia. Statisticians are now more inclined to argue
about the extent to which a method succeeds in solving a data analytic
problem. Much statistical practice revolves around getting
good estimates and standard errors in complicated settings
where statistical uncertainty is smaller than the unquantified
aggregate of many other uncertainties in  scientific
investigation. In such contexts, the
distinction between frequentist and\break Bayesian logic becomes
unimportant and con~temporary practitioners move
freely between frequentist and Bayesian techniques using one or the
other depending on the problem.
Thus, in a review of statistical methods in neurophysiology
in which my colleagues and I discussed both
frequentist and Baye\-sian methods (Kass, Ventura and Brown, \citeyear{KVB2005}),
not only did we not emphasize this dichotomy
but we did not even mention the
distinction between the approaches or
their inferential interpretations.

In fact, in my first publication involving analysis
of neural data (Olson et al., \citeyear{Olsen2001}) we reported
more than a dozen different statistical analyses, some
frequentist, some Bayesian. Furthermore,
methods from the two approaches are sometimes
glued together in a single analysis. For example,
to examine
several neural firing-rate intensity functions
$\lambda^1(t),\ldots,\lambda^p(t)$, assumed to be
smooth functions of time $t$, Behseta et al. (\citeyear{BKMO2007})
developed a frequentist approach to testing the hypothesis
$H_0\dvtx\lambda^1(t)=\cdots=\lambda^p(t)$, for all $t$,
that incorporated BARS smoothing.
Such hybrids are not uncommon, and they
do not force a
practitioner to walk around with mutually inconsistent
interpretations of statistical inference. Figure \ref{fig1} provides a
general framework that encompasses both of the major approaches
to methodology while emphasizing the
inherent gap between data and modeling assumptions, a gap that is
bridged through subjunctive statements.  The advantage of the pragmatic
framework is that it considers frequentist and Bayesian inference to
be equally respectable and allows us to have a consistent
interpretation, without feeling as if we must have split personalities
in order to be competent statisticians.  More to the point, this
framework seems to me to resemble more closely what we do in practice:
statisticians offer inferences couched in a cautionary attitude.
Perhaps we might even say that most practitioners are subjunctivists.

I have emphasized subjunctive statements partly because, on the frequentist
side, they
eliminate any need for long-run interpretation. For
Bayesian methods they eliminate reliance on subjectivism.
The Bayesian point of view was articulated admirably by Jeffreys
(see Robert, Chopin and Rousseau, \citeyear{RCR2010}, and accompanying discussion)
but it became clear, especially from the arguments of Savage
and subsequent investigations in the 1970s,
that the only solid foundation for Bayesianism is subjective
(see Kass and Wasserman,~\citeyear{KW1996},
and Kass, \citeyear{Kass2006}).
Statistical pragmatism pulls us out of that
solipsistic quagmire. On the other hand, I do not mean to
imply that it really does not matter what approach is taken
in a particular instance.
Current attention frequently focuses on challenging, high-dimensional
datasets where
frequentist and Bayesian methods may differ. Statistical
pragmatism is agnostic on this. Instead, procedures should
be judged according to their performance under theoretical
conditions thought to capture relevant real-world variation
in a~particular applied setting. This is where our juxtaposition
of the theoretical world with the real world earns its keep.

I called the story about statistical inference told by
Figure \ref{fig2} ``standard''
because it is imbedded in many introductory texts,
such as the
path-breaking book by Freedman, Pisani and Purves (\citeyear{FPP2007})
and the excellent and very popular
book by Moore and McCabe (\citeyear{MM2005}).
My criticism
is that the standard story misrepresents the way statistical inference
is commonly understood by trained statisticians, portraying it as
analogous to simple random sampling from a finite population.
As I noted, the population versus sampling terminology comes from
Fisher, but I believe the conception in Figure \ref{fig1} is
closer to Fi\-sher's conception of the relationship between
theory and data.
Fisher spoke pointedly of a \textit{hypothetical}
infinite population, but in the standard story of Figure \ref{fig2}
the ``hypothetical'' part of this
notion---which is crucial to the concept---gets
dropped (confer also Lenhard, \citeyear{Lenhard2006}).
I understand Fisher's
``hypothetical'' to connote
what I have here called ``theoretical.''
Fisher
did not anticipate the co-option of his framework and was, in large
part for this reason,  horrified
by subsequent developments by Neyman and Pearson.
The terminology ``theoretical'' avoids this confusion
and thus may offer a clearer representation
of Fisher's idea.\footnote{Fisher
also introduced populations partly because
he used long-run frequency as a foundation for probability,
which statistical pragmatism considers unnecessary.}

We now recognize Neyman and Pearson to have made permanent,
important contributions to statistical inference through
their introduction of hypothesis testing and confidence.
From today's vantage
point, however, their behavioral
interpretation seems quaint, especially when
represented by their famous dictum,
``We are inclined to think that
as far as a particular hypothesis is concerned, no test based upon the
theory of probability can by itself provide any valuable evidence of
the truth or falsehood of that hypothesis.''
Nonetheless, that interpretation seems to have inspired the attitude
behind Figure \ref{fig2}.
In the extreme, one may be led to insist
that statistical inferences are valid only when
some chance mechanism has generated the data.
The problem with the chance-mechanism conception is that it
applies to a rather small part of the real world, where there is
either actual random sampling or situations
described by statistical or quantum physics.
I believe the chance-mechanism conception errs in
declaring that data are assumed to be random variables, rather than
allowing the gap of Figure \ref{fig1} to be bridged\footnote{Because
  probability is introduced
with the goal of drawing conclusions via
statistical inference, it is, in a philosophical sense,
``instrumental.'' See Glymour (\citeyear{Glymour2001}).}
by
statements such as (\ref{bar2}).
In saying this I
am  trying to listen carefully to the voice
in my head that comes from the late David Freedman
(see Freedman and Ziesel, \citeyear{FZ1988}). I imagine he
might call crossing this bridge, in the absence of
an explicit chance mechanism, a leap of faith. In a
strict sense
I am inclined to agree. It seems to me, however, that it
is precisely this leap of faith that
makes statistical reasoning possible in the vast majority of
applications.

Statistical models that go beyond chance mechanisms have been central
to statistical inference since Fisher and Jeffreys, and their role in
reasoning has been considered by many authors (e.g., Cox,
\citeyear{Cox1990};
Lehmann, \citeyear{Lehmann1990}).  An outstanding issue is the extent to which
statistical models are like the theoretical models used throughout
science (see Stanford, \citeyear{Standford2006}).  I would argue, on the one hand, that
they are similar: the most fundamental
belief of any scientist is that the
theoretical and real worlds are aligned.  On the other hand, as
observed in Section \ref{sec2}, statistics is unique in having to face the gap
between theoretical and real worlds every time a model is applied and,
it seems to me, this is a big part of what we offer our scientific
collaborators.  Statistical pragmatism recognizes that all forms of
statistical inference make assumptions, assumptions which can only be
tested very crudely (with such things as goodness-of-fit methods) and
can almost never be verified.  This is not only at the heart of
statistical inference, it is also the great wisdom of our field.

\section*{Acknowledgments}
This work was supported in part by NIH Grant MH064537.
The author is grateful for
comments on an earlier draft
by Brian Junker, Nancy Reid,
Steven Stigler, Larry Wasserman and Gordon Weinberg.

\end{document}